\documentclass[fleqn,usenatbib]{mnras}

\usepackage{newtxtext,newtxmath}

\usepackage[T1]{fontenc}

\DeclareRobustCommand{\VAN}[3]{#2}
\let\VANthebibliography\thebibliography
\def\thebibliography{\DeclareRobustCommand{\VAN}[3]{##3}\VANthebibliography}

\usepackage{graphicx}	
\usepackage{amsmath}	

\usepackage{hyperref} 
\usepackage{orcidlink} 
\usepackage{lineno}

\title[FCN4Flare]{FCN4Flare: Fully Convolution Neural Networks for Flare Detection}

\author[M. Jia et al.]{
Minghui Jia\textsuperscript{\orcidlink{0009-0001-0257-3377}},$^{1,2}$
A-Li Luo\textsuperscript{\orcidlink{0000-0001-7865-2648}}$^{1,2,3}$\thanks{E-mail: lal@nao.cas.cn}
and Bo Qiu$^{4}$
\\
$^{1}$National Astronomical Observatories, Chinese Academy of Sciences, Beijing 100101, China\\
$^{2}$University of Chinese Academy of Sciences, Beijing 100049, China\\
$^{3}$Nanjing college, University of Chinese Academy of Sciences, Nanjing 211135, China\\
$^{4}$University of Science and Technology Beijing, Beijing 100083, China
}

\date{Accepted XXX. Received YYY; in original form ZZZ}

\pubyear{2015}

\begin{document}
\label{firstpage}
\pagerange{\pageref{firstpage}--\pageref{lastpage}}
\maketitle

\begin{abstract}
	Stellar flares offer invaluable insights into stellar magnetic activity and exoplanetary environments. Automated flare detection enables exploiting vast photometric datasets from missions like Kepler. This paper presents FCN4Flare, a deep learning approach using fully convolutional networks (FCN) for precise point-to-point flare prediction regardless of light curve length. Key innovations include the NaN Mask to handle missing data automatedly, and the Mask Dice loss to mitigate severe class imbalance. Experimental results show that FCN4Flare significantly outperforms previous methods, achieving a Dice coefficient of 0.64 compared to the state-of-the-art of 0.12. Applying FCN4Flare to Kepler-LAMOST data, we compile a catalog of 30,285 high-confidence flares across 1426 stars. Flare energies are estimated and stellar/exoplanet properties analyzed, identifying pronounced activity for an M-dwarf hosting a habitable zone planet. This work overcomes limitations of prior flare detection methods via deep learning, enabling new scientific discoveries through analysis of photometric time-series data. Code is available at \href{https://github.com/NAOC-LAMOST/fcn4flare}{https://github.com/NAOC-LAMOST/fcn4flare}
\end{abstract}




\section{Introduction}

Stellar flares may arise from magnetic reconnection events occurring in the stellar atmosphere \citep{reeves_window_2022, doyle_stellar_2018}. They offer important insights into the complex magnetic dynamo processes transpiring deep within stellar cores, as well as the intricate mechanisms orchestrating stellar activity cycles \citep{kowalski_stellar_2024}. By meticulously scrutinizing the distinctive properties of these flares, we gain invaluable insights into the intricate nature of stars themselves. Moreover, beyond their significance in comprehending stellar phenomena, flares emerge as pivotal agents molding the genesis and evolution of exoplanets \cite{poppenhaeger_k_stellar_2015}. The discernment of stellar flare occurrences thus assumes a critical role in our endeavors to assess the potential habitability of exoplanets \citep{atri2021stellar}.

In the contemporary astronomical landscape, the advent of photometric space missions, exemplified by the likes of Kepler \citep{kepler_mission} and TESS \citep{tess_mission}, has ushered in unprecedented opportunities for astronomers to delve into the study of stellar flares. These missions have facilitated the acquisition of vast troves of light curves, presenting us with an exhilarating prospect. However, the sheer magnitude of this data also presents us with a novel challenge: the development of efficient methods for automated flare detection.

To address this challenge, various methodologies, predominantly anchored in outlier detection, have been meticulously crafted \citep{Walkowicz_2011, Hawley_2014, Davenport_2016, Gao_2016, Silverberg_2016, Van_Doorsselaere_2017, Yang_2017, vida2018finding, Yang_2019}. These methods adroitly follow a systematic three-step procedure:
\begin{enumerate}
    \item[(1)] In the initial stage, a meticulous fitting of the quiescent flux or baseline is accomplished through the application of a prewhitening procedure, complemented by an iteratively smooth filter. The combined effect of these techniques is to deftly eliminate extraneous noise and spurious data points, thus laying the foundation for a pristine analysis.
    \item[(2)] With the quiescent flux duly established, stringent criteria, encompassing amplitude, profile, and duration, are selecting the candidate outliers. These selection criteria typically involve setting a threshold on the amplitude, mandating an impulsive surge followed by an exponential decay in the flare profile, and considering the temporal extent of the event.
    \item[(3)] The results are validated using mathematical tests or simulations to establish selection criteria and assess errors. 
\end{enumerate}

To provide concrete instances, \citet{Hawley_2014} adroitly employed an iterative light curve detrending approach to effectively segregate intrinsic variability from potential flares. Flare candidates were duly identified as instances where consecutive points surpassed 2.5 $\sigma$ above the local mean flux for a sustained period. Subsequently, manual confirmation of flares ensued to ensure precision. In a similar vein, \citet{Van_Doorsselaere_2017} meticulously employed thresholds governing the intensity increase and duration of the flares, stipulating a minimum of 3 consecutive points exhibiting an intensity above 3$\sigma$, accompanied by a rapid flux augmentation. Flares, once again, were judiciously validated through manual inspection. \citet{Davenport_2016} adeptly resorted to iterative light curve detrending to model the quiescent brightness. Subsequently, flare candidates were discerningly extracted as outliers from the quiescent model, courtesy of a matched filter approach featuring an analytical flare template. \citet{Gao_2016}, meanwhile, adroitly detrended light curves via the proficient application of Fourier series fitting, mandating that flare candidates exceed a 3$\sigma$ threshold for a minimum duration of 3 data points.

Undoubtedly, the method's utilization of traditional outlier detection comes with two primary limitations, impeding its broad application in large-scale astronomical data analysis. Firstly, the detection process's inherent complexity necessitates continuous iteration, leading to protracted computational times. Such extended processing periods can hinder timely analyses of extensive datasets, making it challenging to keep up with the ever-expanding volume of astronomical information \citep{flatwm2}. Secondly, to effectively suppress the false positive sample rate, the methods rely on the development of highly detailed and stringent criteria. As observed in the study conducted by \citet{Yang_2019}, the earlier flare catalogs suffered from contamination by various false-positive signals and artifacts. Consequently, more rigorous criteria were introduced by \citet{Yang_2019} to discern flares on Kepler light curves. While stricter evaluation criteria enhance the accuracy of flare detection, they introduce further intricacy to an already involved detection process, exacerbating the challenges faced in analyzing large datasets.

The recent application of deep neural networks in detecting stellar flares has shown considerable promise. Notably, Stella \citep{Stella} and Flatwrm2 \citep{flatwm2} have revealed the effectiveness of using convolutional and recurrent networks to determine whether the center of a fixed-length segment of light curves indicates a flare event. By utilizing a sliding window method, these networks can accurately predict flare occurrences throughout a light curve.

The key advantage of employing deep neural networks lies in their inherent capacity for highly parallelizable computations. This attribute allows for expedient processing of substantial datasets, enabling the rapid and efficient detection of flares on a large scale \citep{krizhevsky2012imagenet}. The ability to analyze copious data points in parallel not only accelerates the identification process but also contributes to the overall scalability of the method. Consequently, these deep learning-based approaches hold great promise in tackling the challenges posed by the vast amounts of astronomical data acquired through modern photometric space missions like Kepler and TESS \citep{Stella, flatwm2}.

However, the precision of these methods still need to be improved. Additionally, their use of a fixed sliding window to detect flares in an entire light curve is not efficient enough for large-scale datasets.

Drawing inspiration from the groundbreaking work of \citet{FCN}, our endeavor is to accomplish point-to-point dense prediction, avoiding the sliding window approach. In this paper, we present a fully convolutional network designed for the purpose of detecting stellar flares—henceforth referred to as FCN4Flare. This approach holds the potential to revolutionize the domain of flare detection, propelling it towards unprecedented levels of accuracy and efficiency.

In the following article, Section \ref{sec:FCN4Flare} will introduce the details of FCN4Flare, Section \ref{sec:performance} will discuss the model performance, Section \ref{sec:application} will explore the practical applications of our approach, and Section \ref{sec:conclusion} is the conclusion.

\section{FCN4Flare}
\label{sec:FCN4Flare}

\begin{figure*}
    \centering
    \includegraphics[width=14cm]{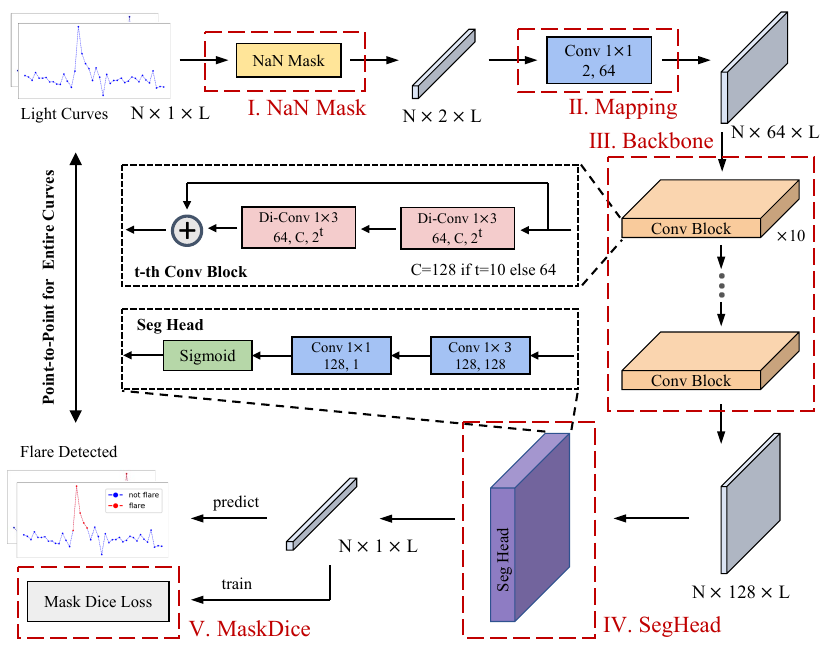}
    \caption{Overview of FCN4Flare. There are 5 components which are framed by red dashed box in the model. NaN Mask deals with NaN in light curves (see section \ref{sec:nanmask}). Mapping is responsible for quickly increasing data dimension so that backbone can extract features. Backbone is stacked by 10 Conv Blocks of which details are shown in the connected black box (see section \ref{sec:backbone}). SegHead of which details is shown in connected black box calculates point-wise probability of flare (see section \ref{sec:seghead}). MaskDice is the loss function designed to solve the problem of extremely imbalance samples in flare detection (see section \ref{sec:maskdice}). Gray cubes represent feature maps and their dimensions are shown beside (e.g., $N \times 2 \times L$ represents $batch\_size \times feature\_dim \times seq\_length$). Blue boxes are convolution layers and pink boxes are dilation convolution layers. Parameters of the boxes are filter size, in channels, out channels and dilation size (only in dilation convolution layers).}
    \label{fig:model_overview}
\end{figure*}

The crux of our task lies in predicting the probability that each flux value in a given light curve belongs to a flare event. To accomplish this, we employ neural networks denoted by $f(X|\Theta)$, where $X=\{x_i \in \mathbb{R} | i=1, 2, \cdots, T\}$ represents the input light curve with a length of $T$. Our objective is to generate a probability vector $P=\{ p_i \in [0, 1] | i=1, 2, \cdots, T \}$, where $p_i$ signifies the likelihood that the flux value $x_i$ is part of a flare event.

A notable feature of our approach is its flexibility regarding input length. Unlike recent methods that utilize equal-length segments of the light curve, generated by sliding a window point by point along the curve, our method can handle light curves of varying lengths. This adaptability allows us to analyze and predict flare probabilities effectively for light curves of different durations, which is particularly advantageous given the inherent variability in astronomical data.

By leveraging the power of neural networks and their capability to discern intricate patterns and relationships within data, our method aims to yield accurate and reliable predictions regarding the likelihood of flare events at individual flux values. With the ability to process light curves of diverse lengths, we anticipate that our approach will contribute significantly to the efficient analysis of large-scale astronomical datasets, unlocking new possibilities for the exploration and understanding of stellar phenomena.

\subsection{Architecture Overview}
\label{sec:architecture}

As depicted in Figure \ref{fig:model_overview}, FCN4Flare is composed of five essential components: NaN Mask, Mapping, Backbone, SegHead, and MaskDice. Each of these components plays a crucial role in the process of flare detection and optimization of the neural network:
\begin{itemize}
    \item \textbf{NaN Mask} This component serves to guide the model in learning where the flux values are NaN (Not-a-Number) and how to handle these NaN values during the flare detection process. Proper handling of NaN values is crucial for accurate predictions in light curves with missing data. See section \ref{sec:nanmask}.
    \item \textbf{Mapping} The Mapping component uses a simple $1 \times 1$ convolution layer to transform the initial input data into a format that the Backbone can easily work with. Think of it as a way to prepare and organize the data, so the next part of the model can better understand and process it. This step is crucial for ensuring that the model can effectively learn from the data. See Section \ref{sec:seghead}.
    \item \textbf{Backbone} The Backbone is the primary component responsible for feature extraction. It is formed by stacking ten Conv Blocks, each contributing to the extraction of relevant features from the input normalized flux sequences. See Section \ref{sec:backbone}.
    \item \textbf{SegHead} The SegHead is responsible for point-wise flare detection of the light curves, regardless of their variable lengths. This component is essential for generating predictions at each point in the light curve, enabling detailed and precise flare detection. See Section \ref{sec:seghead}.
    \item \textbf{MaskDice} MaskDice serves as the optimization objective function for the neural network. Its design is aimed at addressing the challenge of extremely imbalanced samples, which commonly occurs in astronomical datasets. See Section \ref{sec:maskdice}.
\end{itemize}

The input to FCN4Flare is a batch of entire normalized flux sequences of light curves. The normalization process involves dividing the flux values by the median flux, ensuring consistent and comparable representations.

The model's output consists of point-wise flare probabilities, which have the same shape as the input batch. This design allows FCN4Flare to achieve end-to-end training and point-to-point prediction. This advantage not only facilitates efficient model training but also enables accurate and granular predictions, making FCN4Flare a powerful tool for the comprehensive analysis of large-scale astronomical datasets and the identification of stellar flares with precision.

\subsection{NaN Mask}
\label{sec:nanmask}
NaN values, which can cause abnormal calculation results and disrupt the normal training of neural networks, are common in light curves. Previous studies \citep{flatwm2,Stella} addressed this by simply removing NaN data points before feeding light curves into the model. Post-prediction, any flares distributed on both sides of the NaN timestamps were manually excluded. In contrast, our approach automates this process within the model itself.

We introduce the NaN Mask module to guide the model in learning to distinguish NaN values on its own. This module first replaces all NaN flux values with placeholder values (e.g., zeros). It then creates a concatenated matrix by combining the flux vector with an indicator vector, where `1' denotes NaN and `0' denotes non-NaN values. For instance, a light curve with a flux vector $x=[0.1, NaN, 0.3, NaN, NaN]$ results in a matrix $\hat{x}=[[0.1, 0, 0.3, 0, 0], [0, 1, 0, 1, 1]]$. This method allows the model to discern between valid data points and missing data, preventing NaN values from adversely affecting the training and prediction processes.

An alternative approach is to substitute NaN values with realistic values, such as using a median filter to fill in missing data. While this method can provide a more continuous dataset, it may inadvertently introduce bias by assuming a specific pattern in the data. Our NaN Mask approach avoids this by allowing the model to learn the significance of missing data points without making assumptions about their values. We further explore this in the ablation study (see Section \ref{sec:ablation}).

\subsection{SegHead and Mapping}
\label{sec:seghead}
The variable feature shape inherent in light curves poses a significant obstacle when attempting point-to-point prediction using neural networks. Previous approaches, such as those seen in \cite{Stella} and \cite{flatwm2}, were compelled to resort to a sliding window methodology with a fixed width due to the constraints imposed by fully connected (FC) layers. These FC layers demand input data with fixed shapes, rendering them incompatible with variable-length input curves.

In our quest to overcome this limitation, we introduce SegHead module, which enables us to generate variable-length outputs, thereby overcoming the constraint of fixed input shapes.

Consider a flux sequence with an input shape of $1 \times L$, where $L$ denotes the length of the sequence. Upon feature extraction, the resulting feature shape manifests as $F \times L^{'}$, where the dimension $F$ represents the feature channels, and $L^{'}$ denotes the spatial dimension. It is important to note that $L^{'}$ may differ from the original length $L$ due to the utilization of down-sampling layers, which are strategically designed to expand the receptive field of subsequent convolutional filters.

To achieve precise point-to-point prediction, it is crucial that the shape of the output matches that of the input. In other words, we need to convert the feature channels $F$ to $1$ and the spatial dimension $L^{'}$ to $L$. To facilitate this process and for the sake of ease during prediction, we set $L^{'}$ equal to $L$ during the feature extraction phase (see section \ref{sec:backbone} for further elucidation on this aspect). To achieve the conversion of feature dimensions, we resort to employing a convolutional layer, which adeptly modifies the feature channels without altering the spatial dimension. Specifically, we employ convolution filters of size $1 \times 1$ and set the output channels to $1$. As a result, this convolutional layer adeptly accomplishes weighted summation of different channels for each data point, ultimately yielding probabilities through the application of the sigmoid function.

It is noteworthy that the use of a $1 \times 1$ convolutional layer is equivalent to employing a fully connected (FC) layer for each individual point's feature in mathematical terms. However, this convolutional approach imparts greater flexibility to both the input and output, facilitating the accommodation of variable-length input sequences. 

Additionally, in the context of neural networks, as the depth of features increases, the network gains the ability to capture more global and high-level patterns in the data. However, in the case of flare detection, it is crucial to focus on local and fine-grained information within the light curves. To address this challenge effectively, we introduce an additional convolutional layer with a small filter size before the $1 \times 1$ convolution layer as depicted in Figure \ref{fig:model_overview}. This strategic inclusion of a small filter size convolutional layer serves the purpose of capturing and preserving important local information within the data.

Similarly, the application of a $1 \times 1$ convolution layer (Mapping) before the Backbone serves to rapidly increase feature channels while preserving the spatial dimension, as depicted in Figure \ref{fig:model_overview}. This enables the model to effectively capture and leverage higher-level features without compromising the crucial spatial details necessary for accurate flare detection.

\subsection{Backbone}
\label{sec:backbone}
The Backbone, serving as the primary component for feature extraction, plays a pivotal role in the efficacy of the model. As mentioned in section \ref{sec:seghead}, we need to keep feature spacial size the same during feature extracting. So unlike \citet{Stella}, we use dilated convolution layers \citep{V-net,di-conv} instead of combination of traditional convolution and pooling layers.

Dilated convolution is a powerful convolutional operation widely employed in deep learning models for image processing tasks. In this technique, the filter is dilated by inserting zeros between its values, effectively creating a larger filter with gaps or holes between its elements. This dilation strategy augments the receptive field of the filter without necessitating pooling layers, thereby preserving the feature map size.

Inspired by the insightful work of \citet{TS2Vec}, our backbone architecture consists of ten residual convolutional blocks. Each of these blocks comprises two 1-D convolutional layers, with a dilation parameter. The dilation parameter in our backbone architecture is designed to progressively expand the receptive field of each convolutional block, allowing the model to capture both detailed local features and broader contextual information effectively. We chose an exponentially increasing dilation rate, denoted as $2^t$, where $t$ represents the block index, as illustrated in Figure \ref{fig:model_overview}.

To ensure the robustness of our model, we conducted preliminary experiments with different dilation rates, evaluating their impact on the flare detection task. The results indicated that the precise choice of dilation rate had a negligible effect on the overall model performance metrics. Thus, we selected the current configuration for its simplicity.

\subsection{MaskDice}
\label{sec:maskdice}
We have achieved point-to-point prediction on light curves of variable lengths using the aforementioned four components. However, this solution has given rise to a novel challenge, namely, an exceedingly pronounced sample imbalance \citep{diederik2014adam}. As is well-known, the number of flare points within a light curve is significantly smaller in comparison to the number of no-flare points. This significant difference in the distribution of data may compel our neural networks to favor predictions of the no-flare label for all data points, as such an approach would conveniently yield high accuracy (for instance, in the case of a light curve where the flare point rate is merely $1\%$, a model that merely predicts the no-flare label would attain a seemingly impressive accuracy of $99\%$, although such a model would bear no meaningful significance).

It is important to emphasize that such an issue was not prevalent in previous methods \citep{Stella,flatwm2}, as their input was confined to individual curve segments rather than the entire curve. Consequently, they were able to control the distribution of different segment types within the training set, effectively circumventing this predicament. However, in our particular scenario, we confront the daunting task of addressing the challenge of extremely imbalanced samples, necessitating a more sophisticated and nuanced approach.

We use the MaskDice component shown in Figure \ref{fig:model_overview} to solve the problem. It is a loss function based on Dice loss \citep{V-net,ross2017focal,salehi2017tversky}. Specifically, \citet{ross2017focal} introduced the Generalized Dice Loss, which was designed to handle severe class imbalance by weighing different regions according to the class frequencies. This concept of adjusting the weighting mechanism has inspired our approach to further refine the balance between flare and non-flare samples. Similarly, \citet{salehi2017tversky} introduced Tversky Loss, which extended the concept of Dice Loss by adjusting the penalties for false positives and false negatives differently. The MaskDice approach takes inspiration from these works, extending the flexibility of the loss function to better address the unique imbalance problem present in flare detection.

The Dice loss, specifically tailored to address the significant imbalance between foreground and background pixels, is a widely employed loss function in the domain of image semantic segmentation. This loss is determined through the Dice coefficient, a metric employed to gauge the similarity between distinct sets. A higher Dice coefficient signifies a greater degree of similarity between the sets. The Dice coefficient is computed using the formula shown in Equation \ref{eq:dicecoef}, where $DiceCoef \in [0, 1]$ represents the Dice coefficient, and $X$ and $Y$ represent the respective sets involved.

\begin{equation} \label{eq:dicecoef}
    DiceCoef = \frac{2|X \cap Y|}{|X| + |Y|}
\end{equation}

To ensure efficient computation and prevent potential division by zero, the loss is often calculated using the formulation shown in Equation \ref{eq:diceloss_pr}, which is equivalent to Equation \ref{eq:diceloss_set}. In this context, $N$ represents the length of a light curve, $p_i$ and $r_i$ correspond to the model's prediction and the ground truth for the $i$-th point in the light curve respectively, and $\epsilon$ is a small constant introduced to prevent the denominator of the Dice coefficient from becoming zero.

\begin{equation} \label{eq:diceloss_set}
	DiceLoss = 1 - DiceCoef = 1 - \frac{2|X \cap Y|}{|X| + |Y|}
\end{equation}

\begin{equation} \label{eq:diceloss_pr}
    DiceLoss = 1 - \frac{1}{N} \frac{2 \sum^{N}_{i=1} p_i \cdot r_i + \epsilon}{\sum^{N}_{i=1} p_i + \sum^{N}_{i=1} r_i + \epsilon}
\end{equation}

Although Dice Loss has been effective in handling sample imbalance \citep{V-net}, it may not perform optimally in the context of flare detection when dealing with a very small percentage of flare points (approximately 0.5\% in our dataset). To better understand the issue, let's consider a prediction vector $P=\{ p_i \in [0, 1] | i=1, 2, \cdots, T \}$ and a label vector $R = \{ r_i \in \{ 0, 1 \} | i=1, 2, \cdots, T \}$, where $r_i = 1$ denotes that the timestamp is a flare point, and $r_i = 0$ denotes that it is not. In the ideal scenario, $p_i \rightarrow 0$ (e.g., $p_i = 1e-9$) when $r_i = 0$, and it is similar when $r_i = 1$, ensuring that $\sum_{i=1}^{N} p_i \rightarrow \sum_{i=1}^{N} r_i$, and consequently, $DiceLoss \rightarrow 0$.

However, during the initial stages of training, the values of $p_i$ may not be ideal (e.g., $p_i = 0.1$ when $r_i = 0$). In this situation, $\sum_{i=1}^{N} p_i$ might be much larger than $\sum_{i=1}^{N} r_i$ due to the overwhelming proportion of non-flare points. Consequently, the DiceLoss might not decrease significantly, regardless of the nature of the label vector $R$ for a given curve, making it challenging to optimize the neural network parameters effectively.

In essence, the main reason why Dice Loss does not perform well in this context is the exceedingly large proportion of non-flare timestamps. To address this issue, the paper presents a solution in the form of the Mask Dice Loss, which aims to tackle the sample imbalance problem more effectively and improve the performance of flare detection in light curves.

The Mask Dice Loss function is expressed in Equation \ref{Mask_Dice_Loss}, where $u(x)$ represents a step function, as defined in Equation \ref{step_func}, and $m$ is a hyperparameter. The objective of this function is to address the issue of sample imbalance in the context of flare detection. It achieves this by utilizing a simple step function to mask the timestamps for which the prediction probabilities are less than the threshold $m$. By doing so, the loss function penalizes the model for misclassifying non-flare timestamps, thereby mitigating the numerical advantages associated with the significant sample imbalance.

\begin{small}
\begin{equation} \label{Mask_Dice_Loss}
    Mask \; Dice \; Loss = 1 - \frac{1}{N} \frac{2 \sum^{N}_{i=1} u(p_i - m) p_i \cdot r_i + \epsilon}{\sum^{N}_{i=1} u(p_i - m) p_i + \sum^{N}_{i=1} r_i + \epsilon}
\end{equation}
\end{small}

\begin{equation} \label{step_func}
    u(x) = \left \{
    \begin{array}{ll}
         1, x > 0.5  \\
         0, x \leq 0.5 
    \end{array}
    \right.
\end{equation}

The step function $u(x, m)$ acts as a masking mechanism, identifying the timestamps for which the prediction probabilities ($p_i$) are less than the threshold $m$. These masked timestamps, for which the model can easily make correct predictions, have less impact on improving the overall performance compared to the probabilities near the classification threshold boundary. By selectively masking such timestamps, the model is encouraged to pay more attention to the flare samples, enabling the optimizer to guide the optimization of neural network parameters in the right direction from the initial stages of training, even in the presence of the significant sample imbalance problem. As a result, the Mask Dice Loss function helps improve the performance of the model for flare detection in light curves by mitigating the challenges posed by imbalanced data.

\section{Results and Discussion}
\label{sec:result}

\subsection{Data and Training}
The data used in this study is sourced from the catalog presented by \citet{Yang_2019}, which identifies flares in Kepler's long-cadence light curves. 

The Kepler mission, launched in 2009, monitored over 150,000 stars in a fixed region, providing photometric data with a precision of up to 10 ppm for bright targets (V = 9–10) and 100 ppm for fainter targets (V = 13–14) \citep{VanCleve2009}. The mission’s first phase (K1), which provided the data used by \citet{Yang_2019} for flare detection, covered 17 quarters from 2009 to 2013, offering continuous and precise long-cadence (30-minute interval) observations for nearly 200,000 stars \citep{brown2011kepler}.

The catalog records a total of 162,262 flare events across 3,420 distinct stars on 33,214 quarters of light curves. The method of detecting flares used in \citet{Yang_2019} involves a rigorous and detailed approach to ensure the accuracy and credibility of the catalog. It is a multi-step process that includes the following key steps:
\begin{enumerate}
    \item[(1)] Each observation quarter is divided into blocks at points where there are discontinuities longer than 6 hours. Each block is independently adjusted by removing long-term instrumental trends using a third-order polynomial. A Lomb–Scargle periodogram determines the most significant frequency, and a spline filter fits the baseline. An iterative $\sigma$-clipping method removes outliers to clean the data further.
    \item[(2)] Flare candidates are identified under strict conditions:
    \begin{enumerate}
        \item[(i)] At least three consecutive points must exceed a 3$\sigma$ threshold without interruption within 3 hours, with the flare ending at the last point above 1$\sigma$.
        \item[(ii)] The decay phase of a flare must be longer than its rise phase.
        \item[(iii)] Flares must be observed in at least two different quarters if multiple quarters of data are available.
        \item[(iv)] Flares energy must be greater than $10^{34}$ ergs and last least 2 hours if only single quarter is available.
    \end{enumerate}
    \item[(3)] A contamination check is performed to ensure the purity of the flare signal. This involves spatial analyses to rule out photometric contamination from nearby sources, correlation checks with other instruments to confirm the uniqueness of the events, and comparison with known artifacts to identify false positives. Critical cases undergo a manual review by experts to further validate the findings.
\end{enumerate}

The stringent process can significantly reduce the false positive rate of flare detection. While label errors can not be completely cleared up in the catalog, it still provides a robust basis for effectively training and testing deep learning models. This issue is common in machine learning benchmarks, as highlighted by \citet{northcutt_pervasive_2021}, which shows that even widely used datasets like ImageNet comprises label errors at least 6\% of the validation set. Nevertheless, models trained on ImageNet continue to achieve state-of-the-art performance, demonstrating that the presence of label noise does not necessarily compromise the effectiveness of model training.

We also tried to estimate the proportion of incorrect labels in the dataset, but this is a challenging task. In the case of ImageNet, error rates were derived through extensive crowdsourcing efforts, which required significant human and computational resources. For the flare detection dataset, the task is even more complex. Not only would we need to confirm whether each recorded event is a real flare, but we would also need to inspect every light curve in the catalog to ensure no flare events were missed. This would require a level of manual review and verification that is beyond the scope of this study. Therefore, while we acknowledge that some label errors likely exist, estimating their precise rate is impractical given the current resources.

In addition to label accuracy, another key factor in evaluating the model’s robustness is the range of flare energies represented in the training data. As reported by \citet{Yang_2019}, for A-, F-, and G-type stars, flare energies typically range from $10^{34.5}$ to $10^{36}$ erg. For K-type and M-type stars, which are generally more active, the flare energies range from $10^{33.5}$ to $10^{35}$ erg. These ranges reflect the characteristics of the training set and are relevant for understanding the model's applicability across different stellar types.

The 33,214 light curves described in the catalog were divided into training, validation, and test sets using a 6:2:2 ratio. We train all the models including ours and previous methods on the training set from scratch. Model selection was based on performance in the validation set, and the chosen models were then evaluated on the test set. Comparative performance results are discussed in Section \ref{sec:performance}.

\begin{figure}
    \centering
    \includegraphics[width=8cm]{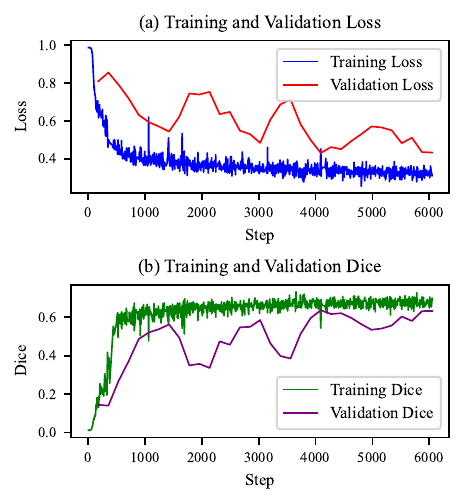}
    \caption{Training status of FCN4Flare.}
    \label{fig:train_loss}
\end{figure}

Our model is trained using two NVIDIA V100 GPUs with a batch size of 150, utilizing the AdamW optimizer at an initial learning rate of $2 \times 10^{-5}$. The training is set to run for a maximum of 50 epochs, incorporating an early stopping mechanism based on validation dice scores. The entire training process takes approximately one hour. Figure \ref{fig:train_loss} illustrates the loss and dice scores during training and evaluation.

\subsection{Detection Performance}
\label{sec:performance}

\begin{figure*}
    \centering
    \includegraphics[width=14cm]{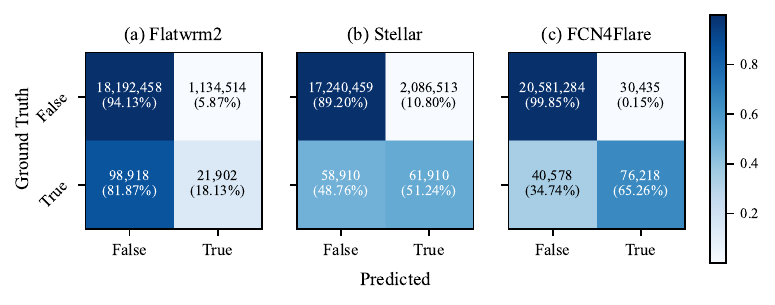}
    \caption{Confusion matrices comparing the performance of Flatwrm2, Stella, and FCN4Flare. Each matrix illustrates the true positive (TP), false positive (FP), true negative (TN), and false negative (FN) counts, with the percentages indicating the proportion of each count relative to the total number of observations. }
    \label{fig:confusion}
\end{figure*}

To rigorously evaluate the efficacy of the FCN4Flare method for stellar flare detection, we employ a suite of statistical metrics that provide a multi-dimensional view of performance. Here are the metrics:

\begin{itemize}
    \item \textbf{Recall} measures the proportion of actual flare events that are correctly identified by the model. A higher recall rate indicates a lower likelihood of missing real flare events. It is calculated as:
    \begin{equation}
        \text{Recall} = \frac{TP}{TP + FN}
    \end{equation}
    where TP and FN correspond true positives and false negatives, respectively.
    
    \item \textbf{Precision} assesses the accuracy of the positive predictions by the model. A higher precision rate indicates that the flares detected are indeed true flares, thereby minimizing the time and resources spent on following up false leads. It is calculated as:
    \begin{equation}
        \text{Precision} = \frac{TP}{TP + FP}
    \end{equation}
    where FP corresponds false positives.
    
    \item \textbf{F1} is the harmonic mean of recall and precision, providing a single measure to balance both metrics. It is calculated as:
    \begin{equation}
        \text{F1} = 2 \times \frac{Recall \times Precision}{Recall + Precision}
    \end{equation}
    
    \item \textbf{AP} is the average precision scores at each threshold. Each precision score is weighted by the increase in recall compared to the previous threshold. AP provides a comprehensive measure that captures the performance across all thresholds which we decide the probability as the flare or not. It is calculate as:
    \begin{equation}
        \text{AP} = \sum (R_n - R_{n-1}) \times P_n
    \end{equation}
    where $R_n$ and $P_n$ represent precision and recall at n-th threshold, respectively.
    
    \item \textbf{AUC-ROC} is the area under the curve plotted with true positive rate (TPR) against false positive rate (FPR) at various threshold settings. It tells us how likely the model is to correctly identify a positive case as being more likely than a negative one when picking randomly. It is calculated as:
    \begin{equation}
        \text{AUC-ROC} = \int_0^1 TPR(x) \, dx
    \end{equation}
    where $x$ corresponds to varying FPR from 0 to 1.
    
    \item \textbf{IoU} also known as the Jaccard index, is the intersection of union. It provides a straightforward metric that quantifies how much the predicted set covers actual set while penalizing false positives and false negatives. It calculated by:
    \begin{equation}
        \text{IoU} = \frac{|A \cap B|}{|A \cup B|}
    \end{equation}
    where A and B are the predicted and actual labels set.
    
    \item \textbf{Dice} coefficient is similar with IoU, but it calculates the size of the overlap times two divided by the sum of the sizes of both sets. It is more sensitive to the size of the intersection relative to the total size of each individual set. It is calculated as:
    \begin{equation}
        \text{Dice} = \frac{2 \times |A \cap B|}{|A| + |B|}
    \end{equation}
\end{itemize}

\begin{table}
\centering
\caption{Performance Metrics for Flare Detection. Entries in bold indicate the highest scores achieved across the models. Note, both Flatwrm2 and Stella have been replicated and trained on our Kepler dataset.}
\label{tab:performance}
\begin{tabular}{cccc}
\hline \hline
            & Flatwrm2  & Stella    & FCN4Flare         \\ \hline
Recall      & 0.26      & 0.50      & \textbf{0.67}     \\ 
Precision   & 0.08      & 0.09      & \textbf{0.69}     \\
F1          & 0.07      & 0.12      & \textbf{0.64}     \\
AP          & 0.02      & 0.03      & \textbf{0.55}     \\
AUC         & 0.61      & 0.68      & \textbf{0.72}     \\
IoU         & 0.04      & 0.07      & \textbf{0.54}     \\
Dice        & 0.07      & 0.12      & \textbf{0.64}     \\
\hline
\end{tabular}
\end{table}

Table \ref{tab:performance} presents a comparative analysis of FCN4Flare with established flare detection methods using selected performance metrics. FCN4Flare's superiority is evident as it outperforms other models in all metrics, particularly showing significant improvements in precision, AP, IoU, and Dice coefficient. These enhancements suggest greater specificity in flare detection, as demonstrated by fewer false positives. To further highlight the model's predictive accuracy, the confusion matrix is shown in Figure \ref{fig:confusion}. This matrix clearly visualizes the reduction in false positives, providing solid evidence of FCN4Flare's enhanced detection capabilities.

\begin{figure*}
    \centering
    \includegraphics{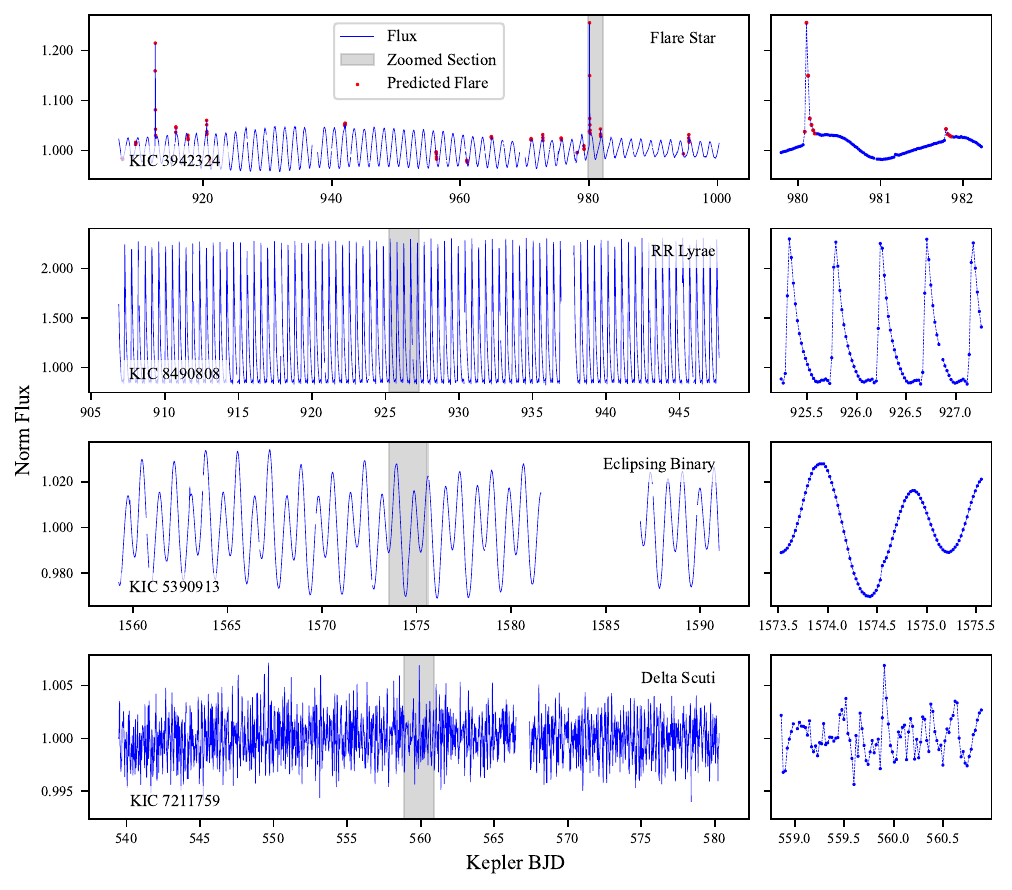}
    \caption{Flare detection results across various star types (typical flare star, RR Lyrae, eclipsing binary, and Delta Scuti), showcasing the precision of our model. The right side zooms into the gray section on the left, highlighting the points predicted as flares (marked in red) by our model. Our model accurately distinguishes true flares from typical flare star, with no incorrect predictions for RR Lyrae, eclipsing binaries, and Delta Scuti which are often misidentified as flare stars by machine learning methods.}
    \label{fig:samples}
\end{figure*}

Additionally, a significant challenge in machine learning for flare detection is that peaks in the light curves of stars with short periods can often be mistaken for flares due to their sharp appearance. This issue is common in various types of stars, including but not limited to RR Lyrae, eclipsing binaries, and Delta Scuti. To demonstrate the high precision of our model, we resampled the light curves of these stars and a typical flare star. The results are displayed in Figure \ref{fig:samples}. None of the samples shown in the figure were included in the training dataset. Flare points detected by our model are marked in red. It is evident that our model does not mistakenly identify flares in these commonly misidentified types. Notably, although these types of light curves were not in the training dataset, our model correctly identifies them, further demonstrating its high precision and robustness.

\subsection{Ablation Study}
\label{sec:ablation}
We present an ablation study to systematically evaluate the contribution of individual components within the FCN4Flare architecture. By sequentially removing key components, we assess their impact on the overall performance of the model, focusing on precision and IoU metrics.

The ablation study involves training variants of the FCN4Flare model, each with a specific component removed or replaced. Each modified version of the model is trained under identical conditions to the full FCN4Flare model to ensure comparability of the results. Performance metrics focusing on precision and IoU are recorded in Table \ref{tab:ablation}. The table displays the metrics scores for each model alongside the percentage change from the full FCN4Flare model, which serves as the baseline.

\begin{table}
    \centering
    \caption{Ablation Study Results for FCN4Flare. The `Replacement' column indicates the method used to handle NaN values or the type of loss function used when a component is removed. `w/o NaN Mask' shows results for removing the NaN Mask and either removing NaNs or using a median filter. `w/o Mask Dice' indicates the use of alternative loss functions.}
    \begin{tabular}{ccll}
        \hline \hline
                        & Replacement         & Precision       & IoU         \\  \hline
        FCN4Flare       & -                   & \textbf{0.69} & \textbf{0.54} \\  \hline
        w/o NaN Mask    & Median Filter       & 0.67 ($\downarrow$\phantom{0}3\%)   & 0.52 ($\downarrow$\phantom{0}4\%)   \\
        w/o NaN Mask    & Remove NaNs         & 0.62 ($\downarrow$10\%)  & 0.46 ($\downarrow$13\%)  \\  \hline
        w/o Mask Dice   & Original Dice Loss  & 0.45 ($\downarrow$35\%)  & 0.39 ($\downarrow$28\%)  \\
        w/o Mask Dice   & Cross-Entropy Loss  & 0.12 ($\downarrow$83\%)  & 0.11 ($\downarrow$80\%)  \\  \hline
        Flatwrm2        & -                   & 0.08          & 0.04          \\
        Stella          & -                   & 0.09          & 0.07          \\  \hline
    \end{tabular}
    \label{tab:ablation}
\end{table}

The baseline model, FCN4Flare, achieves a precision of 0.69 and an IoU of 0.54. When the NaN Mask is removed, two alternative NaN-handling techniques are evaluated: median filtering and direct removal of NaN values. Using a median filter results in a slight reduction in precision (3\%) and IoU (4\%), indicating that this approach is a viable alternative for handling missing data. However, directly removing NaN values, as applied in previous studies such as those by \citet{Stella,flatwm2} and commonly adopted by teams utilizing Kepler and TESS data across various scientific fields \citep{lightkurve,astropy:2013,astropy:2018,astropy:2022,astroquery}, results in a significant performance decline, with a 10\% drop in precision and a 13\% decrease in IoU. This finding indicates that NaN Masking offers a more robust approach for handling missing values in the model.

The impact of the Mask Dice loss is also evaluated by replacing it with alternative loss functions. When the Mask Dice loss is replaced by the original Dice loss, precision and IoU drop by 35\% and 28\%, respectively, indicating that the Mask Dice variant contributes significantly to maintaining model performance. Replacing it with Cross-Entropy loss results in even greater reductions, with precision dropping by 83\% and IoU by 80\%. These findings highlight the Mask Dice loss as a critical component in preserving high accuracy and IoU metrics in the FCN4Flare model.

Finally, the table includes additional models, Flatwrm2 and Stella, which serve as comparative baselines. Both models show considerably lower performance, with precision values of 0.08 and 0.09, and IoU values of 0.04 and 0.07, respectively, underscoring the effectiveness of the FCN4Flare model in combination with the NaN Mask and Mask Dice loss components.

The ablation study provides clear evidence of the importance of NaN Mask and Mask Dice loss components in the FCN4Flare architecture. Each component significantly contributes to the model's overall performance, and their removal or replacement leads to pronounced decreases in both precision and IoU. The table highlights how different strategies for handling NaN values and loss functions affect the model's performance, demonstrating the robustness of the NaN Mask and Mask Dice components.

\section{Application}
\label{sec:application}

To advance our scientific investigation, we deploy FCN4Flare to detect flares on the Kepler-LAMOST dataset, which is generated by cross-matching the Kepler \citep{kepler_mission} DR25 long cadence catalog and LAMOST \citep{cui2012large, luo2012data, luo2015first, zhao2012lamost} DR9 low-resolution catalog with a tolerance of $3^{\prime \prime}$. This dataset encompasses a total of 1,282,142 light curves, representing 87,549 distinct stars.

It is imperative to acknowledge that the data partitioning approach we adopt, utilizing the complete Kepler flare catalog \citep{Yang_2019} for training and a subset of Kepler data for prediction, may be deemed unconventional in machine learning due to concerns of potential data leakage. Nevertheless, in the context of our study, this concern is mitigated. The primary objective of our research is the development of a novel flare detection methodology and the identification of candidates for flare events, rather than a direct comparison of various methods based on specific performance indicators.

The catalog of \citet{Yang_2019} consists exclusively of light curves with flare events, whereas the Kepler-LAMOST cross data set includes many light curves without such events, as it contains a significant number of non-flaring stars. This distinction implies that the majority of light curves in the application dataset may not exhibit detectable flares.

Additionally, the use of the LAMOST cross data set allows for the calculation of spectral parameters. While testing on the entire Kepler dataset could have provided a comprehensive catalog, the inclusion of LAMOST parameters is intended to enhance the scientific value of our results.

In light of these considerations, we have included the outcomes of our model's predictions, which can be valuable for further scientific research and the exploration of flare phenomena.

\subsection{Flare Catalog}

Following the application of our model to the set of light curves, we obtained probabilities assigned to each time step, signifying the likelihood of it belonging to a flare event. The light curves used in this analysis are from the Kepler mission, with a cadence of 30 minutes per point, which provides sufficient temporal resolution for flare detection. A minimum of three consecutive data points is required to confirm a flare event, corresponding to a detection window of 90 minutes.

All light curves from both the \citet{Yang_2019} catalog and the Kepler-LAMOST dataset underwent consistent preprocessing steps, including normalization of flux values by dividing by the median flux, as described in Section \ref{sec:architecture}.

To ensure a higher level of reliability and to minimize the risk of false positive events, we implemented specific screening rules on the prediction results. The rigorous screening process involved the following criteria:

\begin{enumerate}
    \item[(1)] A minimum of three consecutive points with probabilities above 0.5 was required to qualify as a flare event.
    \item[(2)] For these points, the duration from start to peak time must be shorter than from peak to end time, adhering to flares' typical ``rapid rise and slow fall'' pattern.
    \item[(3)] Flux levels before and after the event must be close to each other, indicating a clear deviation from background levels.
    \item[(4)] In the case of a target exhibiting multiple quarters, flares had to occur in at least two quarters to be considered valid. This criterion, following \citet{Yang_2019}, aims to minimize the false positive rate. Approximately 0.2\% of the flare events were removed during this process.
\end{enumerate}

These screening rules were systematically applied during a post-processing step and significantly contributed to enhancing the overall accuracy of our predictions. By carefully filtering the candidate flare events based on these criteria, we could mitigate false detections and ensure the reliability of our flare detection methodology.

\begin{table}
\caption{Parameters of Flare Events. The full table can be accessed at \href{https://github.com/NAOC-LAMOST/fcn4flare/blob/main/catalogs/flare_param.txt}{this URL}.}
\label{tab:flare_param}
\renewcommand{\arraystretch}{1.3}
\resizebox{\linewidth}{!}{
\begin{tabular}{cccccc}
\hline \hline
KIC ID    & Quarter & Start     & End       & Duration & Energy  \\ 
        &         & (days)    & (days)    & (days)   & (erg)    \\
\hline 
7905458 & 10      & 912.7922  & 913.2826  & 0.4904   & 34.2923 \\
7905458 & 10      & 945.2620  & 945.3232  & 0.0613   & 32.6666 \\
7905458 & 10      & 981.6747  & 981.8177  & 0.1430   & 32.8783 \\
7905458 & 10      & 988.1316  & 988.3359  & 0.2043   & 33.6564 \\
7905458 & 16      & 1490.0170 & 1490.1600 & 0.1431   & 33.3342 \\
7905458 & 16      & 1544.3500 & 1544.6970 & 0.3474   & 34.2477 \\
\hline
\end{tabular}}
\end{table}

\begin{table*}
\centering
\caption{Parameters of Flare Stars. The full table can be accessed at \href{https://github.com/NAOC-LAMOST/fcn4flare/blob/main/catalogs/flare_star_param.txt}{this URL}.}
\label{tab:star_param}
\renewcommand{\arraystretch}{1.2}
\resizebox{\linewidth}{!}{
\begin{tabular}{ccccccrccccc}
\hline \hline
KIC ID   & Type & Teff & log g                   & [Fe/H] & Radius & Period & $T_{obs}$   & $T_{fl}/T_{obs}$ & $E_{max}$   & $N_{flare}$ & $L_{fl}/L_{bol}$      \\
    &  & (K)  & $(cm \cdot s^{-2})$ &  & ($R_{\sun}$)  & (day)    & (day)    & (\%)  & (erg)  &   &          \\ \hline
7905458  & G    & 4842 & -0.638                  & 4.353     & 1.065  & 17.642 & 1459.489 & 0.592    & 36.245 & 73     & -4.333 \\
11495571 & dM   & 3523 & 0.57                    & 4.071     & 1.282  & 9.706  & 1470.463 & 0.862    & 36.173 & 158    & -3.958 \\
10120296 & G    & 5490 & -0.219                  & 4.369     & 1.109  & 3.804  & 1459.489 & 0.305    & 36.282 & 40     & -4.237 \\
2852961  & G    & 4722 & -0.34                   & 2.877     & 7.274  & 31.755 & 1470.462 & 0.185    & 37.801 & 14     & -4.465 \\
11970692 & K    & 4707 & -0.246                  & 3.193     & 5.005  & 4.044  & 1470.463 & 0.685    & 37.535 & 73     & -4.321 \\
10330818 & K    & 4344 & -1.018                  & 4.411     & 0.893  & 6.933  & 1470.463 & 0.378    & 35.290 & 69     & -4.652 \\ \hline
\end{tabular}}
\end{table*}

In our investigation, we successfully identified a total of 30,285 distinct flare events occurring in 1426 stars. For each of these events, Table \ref{tab:flare_param} furnishes crucial details, including the start time, end time, and energy estimation.

To estimate the energy associated with each flare event, we adopted a well-established approach based on integrating normalized flux, as described in \citet{chang2017lamost} and \citet{lu2019magnetic}. The Kepler mission provides both uncorrected simple aperture photometry (SAP) flux and presearch data conditioning (PDC) flux, where the instrumental effects of PDC flux have been removed \citep{lu2019magnetic}. In this study, we utilized the PDC flux for the energy estimation. Initially, we normalized the PDC flux by dividing it by the median value. Next, we employed linear interpolation to determine the start and end times of the flare, using the stellar background flux during the flare as reference. These measures helped create a suitable baseline for the estimation process. Subsequently, the flare energy is computed by this equation:
\begin{equation}
    E_{flare} = 4 \pi R_{*}^2 \sigma_{sb} T^4 \int F_{flare} dt
\end{equation}
where $R_{*}$ and $T$ represent the stellar radius and effective temperature, respectively, both of which were sourced from the Kepler Catalog. The symbol $\sigma_{sb}$ denotes the Stefan-Boltzmann constant, while $F_{flare}$ corresponds to the difference between the normalized flux and the interpolated background flux. The energy estimation is expressed in erg units.

\begin{figure}
    \centering
    \includegraphics{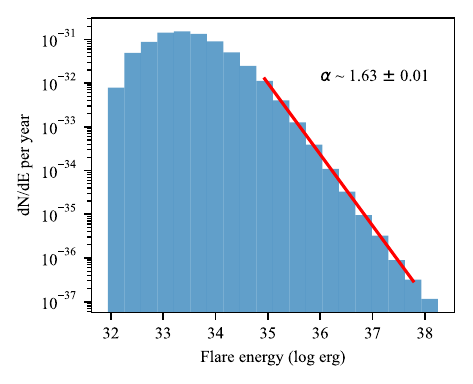}
    \caption{The energy estimation and the frequency distribution of flares. The red solid line is fitted with the slope about 1.63. }
    \label{fig:ffd}
\end{figure}

Figure \ref{fig:ffd} illustrates the energy estimation and the frequency distribution of flares, which intriguingly conform to a power-law relation, expressed as $dN/dE \propto E^{\alpha}$. We apply an ordinary least-square to fit the solid line, whose slope is the $\alpha$ index. This result is presented for reference, aligning with previous studies such as \citet{Shibata2013}, which reported similar power-law exponents in the range of approximately 2 for flare events.

Except for the parameters of the flare events, the parameters of the 1426 flare stars are also listed in Table \ref{tab:star_param}, in which the spectral type is taken from the LAMOST catalog, Teff, log g, [Fe/H] and radius are all taken from the Kepler catalog, and the rotation period is estimated by the Lomb–Scargle periodogram \citep{lomb1976least, scargle1982studies} algorithm, $T_{obs}$ represents the observation time of Kepler, $T_{fl}/T_{obs}$ represents the proportion of the flare duration, $E_max$ represents the maximum detected flare energy, $N_{flare}$ represents the number of detected flare events, and $L_{fl}/L_{bol}$ represents the flare activity, which is calculated as the ratio of the total flare energy to the bolometric luminosity (i.e., $\sum E_{flare} / \int L_{bol} dt = L_{flare} / L_{bol}$) \citep{Yang_2019}.

\begin{table}
\centering
\caption{Incidence of flare stars stratified by spectral types.}
\label{tab:incidence}
\begin{tabular}{crrcc}
\hline  \hline
Type & Nstar & Nflare & Incidence (\%) & Yang (\%) \\ \hline
A    & 2908  & 27     & 0.93           & 1.16 \\
F    & 26642 & 230    & 0.86           & 0.69 \\
G    & 44936 & 648    & 1.44           & 1.46 \\
K    & 10491 & 337    & 3.21           & 2.96 \\
M    & 2346  & 184    & 7.84           & 9.74 \\ \hline
dM   & 837   & 180    & 21.51          & -    \\
gM   & 1509  & 4      & 0.27           & -    \\ \hline
\end{tabular}
\end{table}

Subsequently, we computed the incidence of flare stars stratified by their spectral types, and the findings are presented in Table \ref{tab:incidence}. To further encompass the intrinsic properties of stars, we deliberately differentiated between M-type dwarfs (dM) and M-type giants (gM). Intriguingly, our analysis revealed a pronouncedly elevated incidence of flare stars among dM, while the occurrence of flare events in gM stars was found to be nearly negligible, a trend consistent with observational findings \citep{dzombeta2019flare}. This observation aligns with the current understanding of the underlying flare mechanisms in M-type stars, where M-type dwarfs, characterized by rapid rotation and convective envelopes, exhibit higher magnetic activity, leading to frequent and energetic flaring events \citep{henry2024character,popinchalk2021evaluating,korhonen2013surface}.

In contrast, M-type giants, which have slower rotation rates and more stable internal structures, show much weaker magnetic fields and consequently, almost no detectable flaring activity \citep{sarafopoulos2020dynamo,berdyugina2005starspots,Yang_2019}. These differences in magnetic activity and flaring behavior are attributed to the varying degrees of differential rotation and the presence of convective zones in the stellar interior \citep{dzombeta2019flare,henry2024character,popinchalk2021evaluating,chugainov1991spots}.

\subsection{Flare activity distribution}

\begin{figure*}
    \centering
    \includegraphics{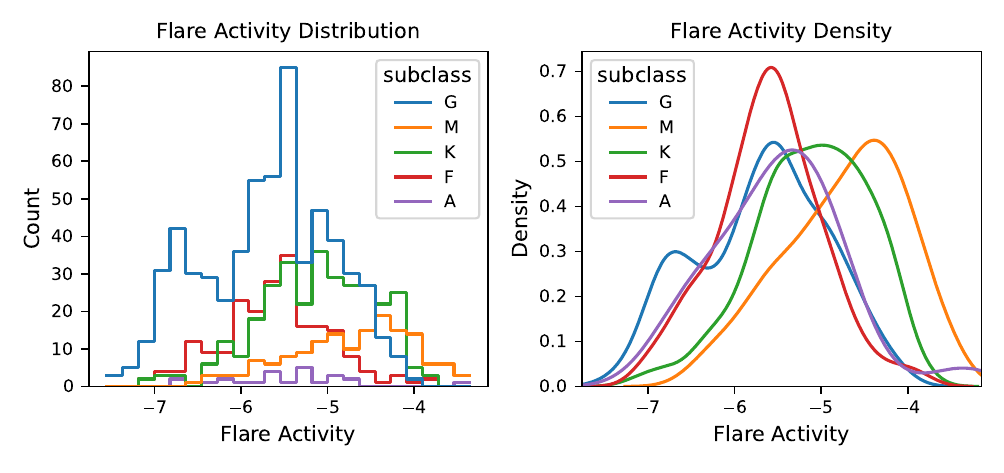}
    \caption{Flare activity distribution. The left panel shows the flare activity distribution of different stellar types, while the right panel is a normalized probabilistic density of flare activity within each stellar type.}
    \label{fig:flare_activity}
\end{figure*}

In the quest to elucidate flare activity across varied stellar spectral types, two quintessential diagrams have been proffered, delineated in Figure \ref{fig:flare_activity}. The first panel presents a histogram detailing the flare activity distribution of different stellar types. Subsequently, the secondary illustration, rendered as a density plot, provides a normalized perspective, artfully portraying the probabilistic density of flare activity within each stellar type. This normalization ensures an equivalency in the integral area beneath each curve, thereby facilitating an unambiguous juxtaposition amongst the subclasses. It is noteworthy that the abscissa of both depictions adopts a logarithmic scale.

A significant observation from these distributions is that the peak of flare activity gradually shifts from M-type stars to F-type stars. Unexpectedly, A-type stars exhibit a level of flare activity that surpasses both F and G types, despite being traditionally regarded as inactive in terms of magnetic activity. This anomaly prompts several possible explanations. One possibility is that the sample of A-type stars with detectable flares is small, rendering these results statistically uncertain. Another plausible explanation is observational bias, as discussed by \citet{Yang_2019}, who reported an unexpectedly high incidence of flaring among A-type stars in specific datasets from the Kepler mission. Their findings suggest that A-type flaring activity, while atypical, could be more common than previously believed and therefore warrants further investigation.

A third possible explanation is that the mechanisms driving flares in A-type stars might differ from those in cooler stars \citep{Yang_2019}. Typically, stellar flares result from magnetic reconnection processes in the chromosphere, a region that is weak in A-type stars as indicated by their low X-ray emissions \citep{balona2013activity,schroder2007x}. However, studies by \citet{balona2013activity,Yang_2019} indicate that flares can indeed occur in A-type stars, suggesting the potential involvement of non-standard magnetic phenomena or interactions with binary companions. This interpretation remains speculative, but it suggests potential challenges to traditional stellar activity models.

\subsection{Chromospheric Dynamics and Stellar Eruptions}

\begin{figure*}
\centering
\includegraphics{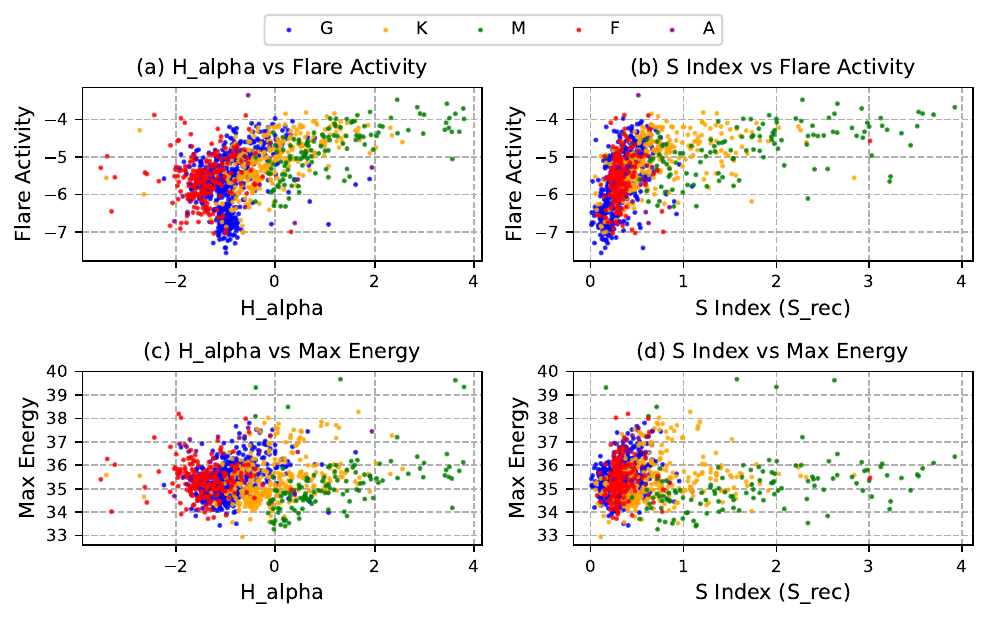}
\caption{Equivalent Widths (EWs) of Spectral Lines in Relation to Stellar Eruption Patterns}
\label{fig:ew_activity}
\end{figure*}

The intricate interplay between chromospheric dynamics and stellar flares has been a subject of sustained academic enquiry. Among the myriad indicators of such dynamism, the $H_\alpha$ equivalent width (EW) and the S Index emerge as particularly salient. Leveraging the LAMOST spectral data, we have discerned the $H_\alpha$ EW and the S Index for our stellar catalog, subsequently elucidating their correlation with stellar flare proclivities. We calculated the $H_\alpha$ equivalent width (EW) using the method outlined in \citet{lu2019magnetic} and the S Index following the approach described in \citet{zhang_stellar_2022}. It is important to note that while the S-index is generally less applicable to A-type stars due to their unique chromospheric properties \citep{shen_s-index_2023}, a small portion of our dataset includes A-type stars, which may not significantly impact the overall findings. These findings are delineated in Figure \ref{fig:ew_activity}. Gleanings from the figure encompass:

\begin{itemize}
    \item \textbf{Interplay of $H_\alpha$ and Flare:} An overwhelming majority of the stellar types, with an especial emphasis on M-type stars, exhibit a pronounced trend wherein an uptick in the $H_\alpha$ EW directly corresponds to an escalation in flare activity. This observation is congruent with antecedent investigations underscoring the nexus between elevated chromospheric dynamism and augmented flare manifestations, particularly within this stellar classification.
    \item \textbf{Correlation between S Index and Flare:} Analogous to the patterns discerned in $H_\alpha$ EW, the S Index also evinces a discernible positive association with flare activities. Notably, the symbiotic relationship for G-type and F-type stars is more palpable in the S Index than in the $H_\alpha$ EW.
    \item \textbf{Nuances between $H_\alpha$ and Max Flare Energies:} The relationship between $H_\alpha$ EW and the max flare energy presents a nuanced scenario, challenging extant paradigms. It appears that elevated $H_\alpha$ EW metrics do not unerringly predicate the magnitude of flares, hinting at the presence of latent variables modulating flare energies. 
    \item \textbf{Interrelation of S Index and Peak Flare Energies:} Mirroring the dynamics between S Index and stellar eruptions, the correlation between the S Index and the apogee of flare energies appears elusive, suggesting a potentially multifaceted interrelation between chromospheric dynamism and peak flare intensities.
\end{itemize}

It is imperative to underscore that the paucity of flare-star specimens in our compilation, especially within the Type A stars, could engender observational biases in discerning correlations.

\begin{table*}
    \caption{Summary of Habitability Indicators for Kepler Objects of Interest: The table lists key characteristics of detected exoplanets, including Kepler ID (`kepid'), Kepler Objects of Interest name (`kepoi name'), official Kepler name (`kepler name'), spectral subclass, radius, effective temperature (`teff'), metallicity (`feh'), surface gravity (`logg'), orbital period, and flare activity. The star with Kepler ID 5938970, situated in the inner habitable zone and exhibiting pronounced magnetic activity, is of particular interest in this study.}
    \label{tab:habitable}
    \centering
    \begin{tabular}{cccccccccc}
    \hline \hline
    kepid    & kepoi name & kepler name  & subclass & radius & teff & feh & logg & period   & flare activity \\
    \hline
    6184894  & K05245.01   & Kepler-1627 b & G        & 0.906       & 5221      & -0.699   & 4.511     & 2.554303 & -4.88317        \\
    11187436 & K01804.01   & Kepler-957 b  & K        & 0.784       & 4813      & -0.419   & 4.575     & 4.648784 & -5.09563        \\
    \textbf{5938970}  & \textbf{K04016.01}   & \textbf{Kepler-1540 b} & \textbf{K}        & \textbf{0.658}       & \textbf{4224}      & \textbf{-0.187}   & \textbf{4.588}     & \textbf{15.94325} & \textbf{-5.15905} \\
    10717220 & K06228.01   & Kepler-1644 b & G        & 0.898       & 5489      & -0.367   & 4.539     & 1.442722 & -5.20815        \\
    10975146 & K01300.01   & Kepler-808 b  & K        & 0.772       & 4369      & -0.978   & 4.513     & 23.90807 & -5.40675        \\
    7100673  & K04032.01   & Kepler-1542 b & F        & 1.331       & 5500      & -0.256   & 4.226     & 0.130462 & -5.48861        \\
    7100673  & K04032.02   & Kepler-1542 c & F        & 1.331       & 5500      & -0.256   & 4.226     & 0.130462 & -5.48861        \\
    7100673  & K04032.03   & Kepler-1542 d & F        & 1.331       & 5500      & -0.256   & 4.226     & 0.130462 & -5.48861        \\
    7100673  & K04032.04   & Kepler-1542 e & F        & 1.331       & 5500      & -0.256   & 4.226     & 0.130462 & -5.48861        \\
    10619192 & K00203.01   & Kepler-17 b   & G        & 0.97        & 5634      & 0.041    & 4.486     & 12.87365 & -5.74915        \\
    5796675  & K00652.01   & Kepler-636 b  & G        & 0.566       & 4628      & -1.447   & 4.791     & 17.64172 & -5.81031        \\
    11551692 & K01781.01   & Kepler-411 c  & K        & 1.993       & 4720      & 0.217    & 3.906     & 12.13283 & -5.82424        \\
    11551692 & K01781.02   & Kepler-411 b  & K        & 1.993       & 4720      & 0.217    & 3.906     & 12.13283 & -5.82424        \\
    11551692 & K01781.03   & Kepler-411 d  & K        & 1.993       & 4720      & 0.217    & 3.906     & 12.13283 & -5.82424        \\
    11554435 & K00063.01   & Kepler-63 b   & G        & 1.065       & 5533      & 0.124    & 4.404     & 5.392375 & -5.8637         \\
    10000941 & K04146.01   & Kepler-1558 b & K        & 1.057       & 4946      & 0.075    & 4.37      & 13.94634 & -5.98847        \\
    11359879 & K00128.01   & Kepler-15 b   & F        & 1.426       & 5718      & 0.362    & 4.181     & 4.922262 & -7.03136        \\
    5357901  & K00188.01   & Kepler-425 b  & K        & 0.674       & 5087      & 0.255    & 4.73      & 3.80461  & -7.16434        \\
    10905746 & K01725.01   & Kepler-1651 b & dM       &             &           &          &           & 6.066414 &                 \\
    8424002  & K03497.01   & Kepler-1512 b & K        &             & 4452.07   & -0.072   & 4.514     & 8.08862  &                 \\
    10736489 & K07368.01   & Kepler-1974 b & K        & 0.9858      & 5279.15   & 0.162    & 4.546     & 2.696199 &      \\
    \hline              
    \end{tabular}
\end{table*}

\subsection{Habitable Zone Exoplanets}
Planetary bodies represent the sole confirmed repositories for the phenomenon of life as we know it. The irradiative influence from a celestial body's central luminary plays a pivotal role in determining the habitability of any given planetary environment. Consequently, the scientific examination of stellar flares assumes paramount importance in the discrimination and selection of potentially life-bearing planets.

The NASA Kepler mission has facilitated the discovery of an extensive array of planetary candidates, many of which have been corroborated by subsequent observational analysis \citep{kane_catalog_2016}. We have executed a cross-match between our flare star catalog and the cumulative table of Kepler Objects of Interest (KOI) \citep{Batalha_2013, Burke_2014, Rowe_2015, Mullally_2015, Coughlin_2016, Thompson_2018}. The resulting data is delineated in Table \ref{tab:habitable}. Within the table, `kepid' refers to the identification, `kepoi\_name' designates a numerical value utilized to identify and monitor a KOI, `kepler\_name' is the nomenclature assigned to the specific planet, and `subclass' denotes the spectral classification of the star as obtained from LAMOST. Additionally, `radius', `teff', `feh', and `logg' symbolize stellar parameters extracted from the Kepler catalog. The table is methodically organized in accordance with flare activity.

Numerous investigations have focused on the intricate subject of habitable zone exoplanets, contributing to a comprehensive understanding of these celestial bodies \citep{kane_catalog_2016, hill_catalog_2023}. In this work, we execute a meticulous cross-reference between the previous table and the Catalog of Habitable Zone Exoplanets, as assembled by \citet{hill_catalog_2023}. Remarkably, one stellar object situated in the inner habitable zone has emerged from this analysis, and it is distinctively highlighted in Table \ref{tab:habitable}. This stellar entity, identified by the Kepler ID 5938970, manifests indications of pronounced magnetic activity.

In the present manuscript, we delineate an exhaustive investigation of this particular star, detailing the salient characteristics such as flare events, flare energy, flare activity indicators, and other pertinent parameters. We harbor the aspiration that our scholarly contribution will significantly augment and influence future research endeavors directed towards the understanding of the habitability potential of this specific stellar object.

\section{Conclusion}
\label{sec:conclusion}
In this paper, we have presented FCN4Flare, a novel deep learning approach for detecting stellar flares in light curves. By leveraging fully convolutional networks, our model achieves end-to-end training and precise point-to-point flare prediction regardless of input length. This overcomes limitations of prior methods relying on fixed-length input segments. 

Through components like NaN Mask and MaskDice, FCN4Flare handles challenges unique to flare detection, including missing data and class imbalance. Compared to the state-of-the-art methods like Flatwrm2 \citep{flatwm2} and Stella \citep{Stella}, which both rely on fixed-length input segments processed through sliding windows, FCN4Flare provides a substantial enhancement by enabling point-to-point prediction across light curves of varying lengths. This approach boosts flare detection precision and efficiency, achieving a Dice coefficient improvement of approximately 433\% compared to the state-of-the-art. These enhancements underscore FCN4Flare’s robustness and adaptability, marking a significant advancement in the domain of stellar flare detection.

Applying FCN4Flare to Kepler-LAMOST data, we have compiled a catalog of 30,285 flare events across 1426 stars. By estimating flare energies and cross-matching with exoplanets, our analysis provides new insights into stellar magnetic activity and planetary habitability. One noteworthy finding is the identification of pronounced flaring behavior on an M-dwarf hosting a planet within the optimistic habitable zone. 

Overall, this work makes multiple contributions: an innovative deep learning architecture for flare detection and a sizable catalog of high-quality flare predictions. FCN4Flare demonstrates the power of deep learning to unlock new scientific discoveries in astronomical time-series data. 

Key directions for future work include enhancing the model, expanding the stellar sample analyzed, and combining our flare predictions with other astronomical data to answer open questions about stellar physics and planetary system evolution. By overcoming long-standing challenges in flare detection, this research opens new avenues for mining the wealth of photometric data from missions like TESS. The transformative potential of deep learning in astronomy is only beginning to be realized.

\section*{Acknowledgements}
This work is supported by the National Science Foundation of China (Nos. 12261141689).  Guoshoujing Telescope (the Large Sky Area Multi-Object Fiber Spectroscopic Telescope, LAMOST) is a National Major Scientific Project built by the Chinese Academy of Sciences.  
Funding for the project has been provided by the National Development and Reform Commission.  
LAMOST is operated and managed by the National Astronomical Observatories, the Chinese Academy of Sciences.

\section*{Data Availability}

The catalog used for training the model in this study can be accessed at \href{https://content.cld.iop.org/journals/0067-0049/241/2/29/revision1/apjsab0d28t2_mrt.txt}{this URL}. The observational data utilized in this research were obtained from the official \href{https://archive.stsci.edu/missions-and-data/kepler}{Kepler mission website}. Additionally, the code and catalogs referenced in this paper are available at \href{https://github.com/NAOC-LAMOST/fcn4flare}{https://github.com/NAOC-LAMOST/fcn4flare}.

\bibliographystyle{mnras}
\bibliography{references}

\begin{thebibliography}{}
\makeatletter
\relax
\def\mn@urlcharsother{\let\do\@makeother \do\$\do\&\do\#\do\^\do\_\do\%\do\~}
\def\mn@doi{\begingroup\mn@urlcharsother \@ifnextchar [ {\mn@doi@}
  {\mn@doi@[]}}
\def\mn@doi@[#1]#2{\def\@tempa{#1}\ifx\@tempa\@empty \href
  {http://dx.doi.org/#2} {doi:#2}\else \href {http://dx.doi.org/#2} {#1}\fi
  \endgroup}
\def\mn@eprint#1#2{\mn@eprint@#1:#2::\@nil}
\def\mn@eprint@arXiv#1{\href {http://arxiv.org/abs/#1} {{\tt arXiv:#1}}}
\def\mn@eprint@dblp#1{\href {http://dblp.uni-trier.de/rec/bibtex/#1.xml}
  {dblp:#1}}
\def\mn@eprint@#1:#2:#3:#4\@nil{\def\@tempa {#1}\def\@tempb {#2}\def\@tempc
  {#3}\ifx \@tempc \@empty \let \@tempc \@tempb \let \@tempb \@tempa \fi \ifx
  \@tempb \@empty \def\@tempb {arXiv}\fi \@ifundefined
  {mn@eprint@\@tempb}{\@tempb:\@tempc}{\expandafter \expandafter \csname
  mn@eprint@\@tempb\endcsname \expandafter{\@tempc}}}

\bibitem[\protect\citeauthoryear{{Astropy Collaboration} et~al.,}{{Astropy
  Collaboration} et~al.}{2013}]{astropy:2013}
{Astropy Collaboration} et~al., 2013, \mn@doi [\aap]
  {10.1051/0004-6361/201322068}, \href
  {http://adsabs.harvard.edu/abs/2013A%26A...558A..33A} {558, A33}

\bibitem[\protect\citeauthoryear{{Astropy Collaboration} et~al.,}{{Astropy
  Collaboration} et~al.}{2018}]{astropy:2018}
{Astropy Collaboration} et~al., 2018, \mn@doi [\aj] {10.3847/1538-3881/aabc4f},
  \href {https://ui.adsabs.harvard.edu/abs/2018AJ....156..123A} {156, 123}

\bibitem[\protect\citeauthoryear{{Astropy Collaboration} et~al.,}{{Astropy
  Collaboration} et~al.}{2022}]{astropy:2022}
{Astropy Collaboration} et~al., 2022, \mn@doi [\apj]
  {10.3847/1538-4357/ac7c74}, \href
  {https://ui.adsabs.harvard.edu/abs/2022ApJ...935..167A} {935, 167}

\bibitem[\protect\citeauthoryear{Atri \& Mogan}{Atri \&
  Mogan}{2021}]{atri2021stellar}
Atri D.,  Mogan S. R.~C.,  2021, Monthly Notices of the Royal Astronomical
  Society: Letters, 500, L1

\bibitem[\protect\citeauthoryear{Balona}{Balona}{2013}]{balona2013activity}
Balona L.,  2013, Monthly Notices of the Royal Astronomical Society, 431, 2240

\bibitem[\protect\citeauthoryear{Batalha et~al.,}{Batalha
  et~al.}{2013}]{Batalha_2013}
Batalha N.~M.,  et~al., 2013, \mn@doi [The Astrophysical Journal Supplement
  Series] {10.1088/0067-0049/204/2/24}, 204, 24

\bibitem[\protect\citeauthoryear{Berdyugina}{Berdyugina}{2005}]{berdyugina2005starspots}
Berdyugina S.,  2005, Living Reviews in Solar Physics

\bibitem[\protect\citeauthoryear{Borucki et~al.,}{Borucki
  et~al.}{2010}]{kepler_mission}
Borucki W.~J.,  et~al., 2010, \mn@doi [Science] {10.1126/science.1185402}, 327,
  977

\bibitem[\protect\citeauthoryear{Brown, Latham, Everett  \& Esquerdo}{Brown
  et~al.}{2011}]{brown2011kepler}
Brown T.~M.,  Latham D.~W.,  Everett M.~E.,   Esquerdo G.~A.,  2011, The
  Astronomical Journal, 142, 112

\bibitem[\protect\citeauthoryear{Burke et~al.,}{Burke
  et~al.}{2014}]{Burke_2014}
Burke C.~J.,  et~al., 2014, \mn@doi [The Astrophysical Journal Supplement
  Series] {10.1088/0067-0049/210/2/19}, 210, 19

\bibitem[\protect\citeauthoryear{Chang et~al.,}{Chang
  et~al.}{2017}]{chang2017lamost}
Chang H.-Y.,  et~al., 2017, The Astrophysical Journal, 834, 92

\bibitem[\protect\citeauthoryear{Chugainov}{Chugainov}{1991}]{chugainov1991spots}
Chugainov P.,  1991, Astrophysics

\bibitem[\protect\citeauthoryear{Coughlin et~al.,}{Coughlin
  et~al.}{2016}]{Coughlin_2016}
Coughlin J.~L.,  et~al., 2016, \mn@doi [The Astrophysical Journal Supplement
  Series] {10.3847/0067-0049/224/1/12}, 224, 12

\bibitem[\protect\citeauthoryear{Cui et~al.,}{Cui et~al.}{2012}]{cui2012large}
Cui X.-Q.,  et~al., 2012, Research in Astronomy and Astrophysics, 12, 1197

\bibitem[\protect\citeauthoryear{Davenport}{Davenport}{2016}]{Davenport_2016}
Davenport J. R.~A.,  2016, \mn@doi [The Astrophysical Journal]
  {10.3847/0004-637X/829/1/23}, 829, 23

\bibitem[\protect\citeauthoryear{Diederik}{Diederik}{2014}]{diederik2014adam}
Diederik P.~K.,  2014, (No Title)

\bibitem[\protect\citeauthoryear{Doorsselaere, Shariati  \&
  Debosscher}{Doorsselaere et~al.}{2017}]{Van_Doorsselaere_2017}
Doorsselaere T.~V.,  Shariati H.,   Debosscher J.,  2017, \mn@doi [The
  Astrophysical Journal Supplement Series] {10.3847/1538-4365/aa8f9a}, 232, 26

\bibitem[\protect\citeauthoryear{Doyle et~al.,}{Doyle
  et~al.}{2018}]{doyle_stellar_2018}
Doyle J.~G.,  et~al., 2018, \mn@doi [Monthly Notices of the Royal Astronomical
  Society] {10.1093/mnras/sty032}, 475, 2842

\bibitem[\protect\citeauthoryear{Dzombeta \& Percy}{Dzombeta \&
  Percy}{2019}]{dzombeta2019flare}
Dzombeta K.,  Percy J.,  2019, in TSpace. \url
  {https://tspace.library.utoronto.ca/handle/1807/97060}

\bibitem[\protect\citeauthoryear{Feinstein, Montet, Ansdell, Nord, Bean,
  G{\"u}nther, Gully-Santiago  \& Schlieder}{Feinstein et~al.}{2020}]{Stella}
Feinstein A.~D.,  Montet B.~T.,  Ansdell M.,  Nord B.,  Bean J.~L.,
  G{\"u}nther M.~N.,  Gully-Santiago M.,   Schlieder J.~E.,  2020, The
  Astronomical Journal

\bibitem[\protect\citeauthoryear{Gao, Xin, Liu, Zhang  \& Gao}{Gao
  et~al.}{2016}]{Gao_2016}
Gao Q.,  Xin Y.,  Liu J.-F.,  Zhang X.-B.,   Gao S.,  2016, \mn@doi [The
  Astrophysical Journal Supplement Series] {10.3847/0067-0049/224/2/37}, 224,
  37

\bibitem[\protect\citeauthoryear{Ginsburg et~al.,}{Ginsburg
  et~al.}{2019}]{astroquery}
Ginsburg A.,  et~al., 2019, \mn@doi [The Astronomical Journal]
  {10.3847/1538-3881/aafc33}, 157, 98

\bibitem[\protect\citeauthoryear{Hawley, Davenport, Kowalski, Wisniewski, Hebb,
  Deitrick  \& Hilton}{Hawley et~al.}{2014}]{Hawley_2014}
Hawley S.~L.,  Davenport J. R.~A.,  Kowalski A.~F.,  Wisniewski J.~P.,  Hebb
  L.,  Deitrick R.,   Hilton E.~J.,  2014, \mn@doi [The Astrophysical Journal]
  {10.1088/0004-637X/797/2/121}, 797, 121

\bibitem[\protect\citeauthoryear{Henry \& Jao}{Henry \&
  Jao}{2024}]{henry2024character}
Henry T.,  Jao W.,  2024, Annual Review of Astronomy and Astrophysics

\bibitem[\protect\citeauthoryear{Hill, Bott, Dalba, Fetherolf, Kane, Kopparapu,
  Li  \& Ostberg}{Hill et~al.}{2023}]{hill_catalog_2023}
Hill M.~L.,  Bott K.,  Dalba P.~A.,  Fetherolf T.,  Kane S.~R.,  Kopparapu R.,
  Li Z.,   Ostberg C.,  2023, \mn@doi [The Astronomical Journal]
  {10.3847/1538-3881/aca1c0}, 165, 34

\bibitem[\protect\citeauthoryear{Kane et~al.,}{Kane
  et~al.}{2016}]{kane_catalog_2016}
Kane S.~R.,  et~al., 2016, \mn@doi [The Astrophysical Journal]
  {10.3847/0004-637X/830/1/1}, 830, 1

\bibitem[\protect\citeauthoryear{Korhonen}{Korhonen}{2013}]{korhonen2013surface}
Korhonen H.,  2013, in Proceedings of the International Astronomical Union.
  \url
  {https://www.cambridge.org/core/journals/proceedings-of-the-international-astronomical-union/article/surface-magnetism-of-cool-giant-and-supergiant-stars/27BB71D4478EB6EC48953BBA4E3913D0}

\bibitem[\protect\citeauthoryear{Kowalski}{Kowalski}{2024}]{kowalski_stellar_2024}
Kowalski A.~F.,  2024, arXiv preprint arXiv:2402.07885

\bibitem[\protect\citeauthoryear{Krizhevsky, Sutskever  \& Hinton}{Krizhevsky
  et~al.}{2012}]{krizhevsky2012imagenet}
Krizhevsky A.,  Sutskever I.,   Hinton G.~E.,  2012, Advances in neural
  information processing systems, 25

\bibitem[\protect\citeauthoryear{{Lightkurve Collaboration}
  et~al.,}{{Lightkurve Collaboration} et~al.}{2018}]{lightkurve}
{Lightkurve Collaboration} et~al., 2018, {Lightkurve: Kepler and TESS time
  series analysis in Python}, Astrophysics Source Code Library (\mn@eprint
  {ascl} {1812.013})

\bibitem[\protect\citeauthoryear{Lomb}{Lomb}{1976}]{lomb1976least}
Lomb N.~R.,  1976, Astrophysics and space science, 39, 447

\bibitem[\protect\citeauthoryear{Long, Shelhamer  \& Darrell}{Long
  et~al.}{2015}]{FCN}
Long J.,  Shelhamer E.,   Darrell T.,  2015, in Proceedings of the IEEE
  Conference on Computer Vision and Pattern Recognition (CVPR).

\bibitem[\protect\citeauthoryear{Lu, Zhang, Shi, Han, Fan, Long  \& Pi}{Lu
  et~al.}{2019}]{lu2019magnetic}
Lu H.-p.,  Zhang L.-y.,  Shi J.,  Han X.~L.,  Fan D.,  Long L.,   Pi Q.,  2019,
  The Astrophysical Journal Supplement Series, 243, 28

\bibitem[\protect\citeauthoryear{Luo et~al.,}{Luo et~al.}{2012}]{luo2012data}
Luo A.-L.,  et~al., 2012, Research in Astronomy and Astrophysics, 12, 1243

\bibitem[\protect\citeauthoryear{Luo et~al.,}{Luo et~al.}{2015}]{luo2015first}
Luo A.-L.,  et~al., 2015, Research in Astronomy and Astrophysics, 15, 1095

\bibitem[\protect\citeauthoryear{Milletari, Navab  \& Ahmadi}{Milletari
  et~al.}{2016}]{V-net}
Milletari F.,  Navab N.,   Ahmadi S.-A.,  2016, in 2016 fourth international
  conference on 3D vision (3DV). pp 565--571

\bibitem[\protect\citeauthoryear{Mullally et~al.,}{Mullally
  et~al.}{2015}]{Mullally_2015}
Mullally F.,  et~al., 2015, \mn@doi [The Astrophysical Journal Supplement
  Series] {10.1088/0067-0049/217/2/31}, 217, 31

\bibitem[\protect\citeauthoryear{Northcutt, Athalye  \& Mueller}{Northcutt
  et~al.}{2021}]{northcutt_pervasive_2021}
Northcutt C.~G.,  Athalye A.,   Mueller J.,  2021, Pervasive Label Errors in
  Test Sets Destabilize Machine Learning Benchmarks,
  \mn@doi{10.48550/arXiv.2103.14749}, \url {http://arxiv.org/abs/2103.14749}

\bibitem[\protect\citeauthoryear{Popinchalk, Faherty  \& Kiman}{Popinchalk
  et~al.}{2021}]{popinchalk2021evaluating}
Popinchalk M.,  Faherty J.,   Kiman R.,  2021, The Astrophysical Journal, 1538

\bibitem[\protect\citeauthoryear{{Poppenhaeger, K.}}{{Poppenhaeger,
  K.}}{2015}]{poppenhaeger_k_stellar_2015}
{Poppenhaeger, K.} 2015, \mn@doi [EPJ Web of Conferences]
  {10.1051/epjconf/201510105002}, 101

\bibitem[\protect\citeauthoryear{Reeves}{Reeves}{2022}]{reeves_window_2022}
Reeves K.~K.,  2022, \mn@doi [Frontiers in Astronomy and Space Sciences]
  {10.3389/fspas.2022.1041951}, 9

\bibitem[\protect\citeauthoryear{Ricker et~al.,}{Ricker
  et~al.}{2014}]{tess_mission}
Ricker G.~R.,  et~al., 2014, \mn@doi [Journal of Astronomical Telescopes,
  Instruments, and Systems] {10.1117/1.JATIS.1.1.014003}, 1, 014003

\bibitem[\protect\citeauthoryear{Ross \& Doll{\'a}r}{Ross \&
  Doll{\'a}r}{2017}]{ross2017focal}
Ross T.-Y.,  Doll{\'a}r G.,  2017, in proceedings of the IEEE conference on
  computer vision and pattern recognition. pp 2980--2988

\bibitem[\protect\citeauthoryear{Rowe et~al.,}{Rowe et~al.}{2015}]{Rowe_2015}
Rowe J.~F.,  et~al., 2015, \mn@doi [The Astrophysical Journal Supplement
  Series] {10.1088/0067-0049/217/1/16}, 217, 16

\bibitem[\protect\citeauthoryear{Salehi, Erdogmus  \& Gholipour}{Salehi
  et~al.}{2017}]{salehi2017tversky}
Salehi S. S.~M.,  Erdogmus D.,   Gholipour A.,  2017, in International workshop
  on machine learning in medical imaging. pp 379--387

\bibitem[\protect\citeauthoryear{Sarafopoulos}{Sarafopoulos}{2020}]{sarafopoulos2020dynamo}
Sarafopoulos D.,  2020, arXiv preprint

\bibitem[\protect\citeauthoryear{Scargle}{Scargle}{1982}]{scargle1982studies}
Scargle J.~D.,  1982, Astrophysical Journal, Part 1, vol. 263, Dec. 15, 1982,
  p. 835-853., 263, 835

\bibitem[\protect\citeauthoryear{Schr{\"o}der \& Schmitt}{Schr{\"o}der \&
  Schmitt}{2007}]{schroder2007x}
Schr{\"o}der C.,  Schmitt J.,  2007, Astronomy \& Astrophysics, 475, 677

\bibitem[\protect\citeauthoryear{Shen}{Shen}{2023}]{shen_s-index_2023}
Shen Y.-F.,  2023, \mn@doi [Scientific Reports] {10.1038/s41598-023-48590-8},
  13, 21095

\bibitem[\protect\citeauthoryear{Shibata et~al.,}{Shibata
  et~al.}{2013}]{Shibata2013}
Shibata K.,  et~al., 2013, \mn@doi [Publications of the Astronomical Society of
  Japan] {10.1093/pasj/65.3.49}, 65, 49

\bibitem[\protect\citeauthoryear{Silverberg, Kowalski, Davenport, Wisniewski,
  Hawley  \& Hilton}{Silverberg et~al.}{2016}]{Silverberg_2016}
Silverberg S.~M.,  Kowalski A.~F.,  Davenport J. R.~A.,  Wisniewski J.~P.,
  Hawley S.~L.,   Hilton E.~J.,  2016, \mn@doi [The Astrophysical Journal]
  {10.3847/0004-637X/829/2/129}, 829, 129

\bibitem[\protect\citeauthoryear{Thompson et~al.,}{Thompson
  et~al.}{2018}]{Thompson_2018}
Thompson S.~E.,  et~al., 2018, \mn@doi [The Astrophysical Journal Supplement
  Series] {10.3847/1538-4365/aab4f9}, 235, 38

\bibitem[\protect\citeauthoryear{Van~Cleve \& Caldwell}{Van~Cleve \&
  Caldwell}{2009}]{VanCleve2009}
Van~Cleve J.~E.,  Caldwell D.~A.,  2009, Tech. Rep. KSCI-19033, Kepler: A
  Search for Terrestrial Planets, Kepler Instrument Handbook, \url
  {https://archive.stsci.edu/kepler/manuals/KSCI-19033-001.pdf}.
NASA Ames Research Center, Moffett Field, CA, \url
  {https://archive.stsci.edu/kepler/manuals/KSCI-19033-001.pdf}

\bibitem[\protect\citeauthoryear{Vida \& Roettenbacher}{Vida \&
  Roettenbacher}{2018}]{vida2018finding}
Vida K.,  Roettenbacher R.~M.,  2018, Astronomy \& Astrophysics, 616, A163

\bibitem[\protect\citeauthoryear{Vida, B{\'o}di, Szklenar  \& Seli}{Vida
  et~al.}{2021}]{flatwm2}
Vida K.,  B{\'o}di A.,  Szklenar T.,   Seli B.,  2021, Astronomy and
  Astrophysics

\bibitem[\protect\citeauthoryear{Walkowicz et~al.,}{Walkowicz
  et~al.}{2011}]{Walkowicz_2011}
Walkowicz L.~M.,  et~al., 2011, \mn@doi [The Astronomical Journal]
  {10.1088/0004-6256/141/2/50}, 141, 50

\bibitem[\protect\citeauthoryear{Wang, Yang, Yue, Duan, Tong, Xu  \&
  Huang}{Wang et~al.}{2021}]{TS2Vec}
Wang Y.,  Yang T.,  Yue Z.,  Duan J.,  Tong Y.,  Xu B.,   Huang C.,  2021,
  arXiv: Learning

\bibitem[\protect\citeauthoryear{Yang \& Liu}{Yang \& Liu}{2019}]{Yang_2019}
Yang H.,  Liu J.,  2019, \mn@doi [The Astrophysical Journal Supplement Series]
  {10.3847/1538-4365/ab0d28}, 241, 29

\bibitem[\protect\citeauthoryear{Yang et~al.,}{Yang et~al.}{2017}]{Yang_2017}
Yang H.,  et~al., 2017, \mn@doi [The Astrophysical Journal]
  {10.3847/1538-4357/aa8ea2}, 849, 36

\bibitem[\protect\citeauthoryear{Yu \& Koltun}{Yu \& Koltun}{2015}]{di-conv}
Yu F.,  Koltun V.,  2015, arXiv preprint arXiv:1511.07122

\bibitem[\protect\citeauthoryear{Zhang, Zhang, He, Song, Luo  \& Zhang}{Zhang
  et~al.}{2022}]{zhang_stellar_2022}
Zhang W.,  Zhang J.,  He H.,  Song Z.,  Luo A.,   Zhang H.,  2022, Stellar
  {Chromospheric} {Activity} {Database} of {Solar}-like {Stars} {Based} on the
  {LAMOST} {Low}-{Resolution} {Spectroscopic} {Survey},
  \mn@doi{10.48550/arXiv.2209.15255}, \url {http://arxiv.org/abs/2209.15255}

\bibitem[\protect\citeauthoryear{Zhao, Zhao, Chu, Jing  \& Deng}{Zhao
  et~al.}{2012}]{zhao2012lamost}
Zhao G.,  Zhao Y.-H.,  Chu Y.-Q.,  Jing Y.-P.,   Deng L.-C.,  2012, Research in
  Astronomy and Astrophysics, 12, 723

\makeatother
\end{thebibliography}





\bsp	
\label{lastpage}
\end{document}